\documentclass[useAMS,usegraphicx]{mn2e}

\title[Magnetic field variability of gamma Equ (HD 201601) ]
{Periods of magnetic field variations in the Ap star 
   $\gamma$~Equulei (HD 201601) }

\author[V.D. Bychkov, L.V. Bychkova, J. Madej ]{V.D. Bychkov$^{1}$
   \thanks{E-mail: vbych@sao.ru (VDB) }, 
    L.V. Bychkova$^{1}$, J. Madej$^{2}$   \footnotemark[1] \\
$^{1}$ Special Astrophysical Observatory of the Russian
        Academy of Sciences (SAO), Nizhnij Arkhyz, 369167 Russia \\
$^{2}$ Astronomical Observatory, University of Warsaw,
        Al. Ujazdowskie 4, 00-478 Warszawa, Poland  }

\begin{document}
\date{Accepted .................................. }
\pagerange{\pageref{firstpage}--\pageref{lastpage}} \pubyear{2015}

\maketitle
\label{firstpage}

\begin{abstract}
This paper presents a series of 95 new measurements of the longitudinal (effective)
magnetic field $B_e$ of the Ap star $\gamma$ Equ (HD 201601). Observations
were obtained at the coud\'e focus of the 1-m reflector at the Special Astrophysical
Observatory (SAO RAS) in Russia over a time period of 4190 days 
(more than 11 years). We compiled
a long record of $B_e$ points, adding our measurements to all published data.  
The time series of magnetic data consists of 395 $B_e$ points extending for
24488 days, i.e. over 67 years. Various methods of period determination
were examined for the case in which the length of the observed time series
is rather short and amounts only to ~69 percent of the period.
We argue that the fitting of a sine wave to the observed $B_e$ points by
least squares yields the most reliable period in the case of $\gamma$ Equ. 
Therefore, the best period for long-term magnetic variations of $\gamma$ Equ,
and hence the rotational period, is {\rm
$P_{\rm rot}=35462.5 \pm 1149$ days $= 97.16 \pm 3.15$ years. } 
\end{abstract}


\section{Introduction}

Ap star $\gamma$ Equ (HD 201601; HR 8097) is an apparently bright
object ($V=4.7$ mag) with a strong global magnetic field. Its longitudinal
component $B_e$ (effective magnetic field) exhibits very slow variations
plus significant noise in the range from $\approx 1000$ G to $-1600$ G.
The total time period over which the magnetic field $B_e$ data have 
been recorded exceeds 60 years. 

The long-term variability of the longitudinal magnetic field of $\gamma$ Equ
most likely is due to the very slow rotation of the star. That feature has 
been investigated in many papers [see 
Bonsack \& Pilachowski (1974); Leroy et al. (1994); Bychkov \& Shtol (1997);
Scholz et al. (1997); Bychkov et al. (2006); Savanov et al. (2014)]. 
After smoothing, all available $B_e$ points accumulated in published
sources plotted against time are arranged as part of a sine wave with 
a long period. Bychkov et al. (2006) determined the period
of long-term magnetic variations as $P_{\rm mag} = 91.1 \pm 3.6$ years.
However, Savanov et al. (2014) determined this period as
$P_{\rm mag}=93 \pm 3$ years after adding their new $B_e$ measurements.

Such a discrepancy between periods $P_{\rm mag}$ in both papers is probably
the result of the application of different numerical methods in the 
particular case when an unknown period of $B_e$ variations is longer than
the length of the observed time series. We believe that the most commonly 
used methods of period determination fail in such a case and yield 
inaccurate results.

Our paper aims to examine the systematic errors introduced by various
commonly used methods of period determination in the particular case of
the Ap star $\gamma$ Equ. We aim to reliably determine the period of 
long-term $B_e$ variations, as well as to search for 
a possibly more rapid variability of this star.

\section{Observational data}

The longitudinal magnetic field $B_e$ of $\gamma$ Equ was routinely measured
at the Special Astrophysical Observatory (SAO RAS) in Russia over many years.
Observations were carried out at the coud\'e focus of the 1-m reflector.
During the observational period (4190 days $\approx$ 11 years) we acquired
a total of 95 $B_e$ measurements for this object. 
Those observations are shown in Table~\ref{tab:saores} of Appendix A.

\subsection{Details of our observations}

The magnetic measurements of Gamma Equ were derived from the Zeeman spectra, 
obtained in the coud\'e focus of the 1-m telescope of SAO RAS. The instrumentation
and data reduction procedures were described in Bychkov (2008). In our case,
we analysed high-quality spectral data ($R = 45000$, CCD, S/N $\approx 100$)
and processed them using  the MIDAS software. The method for obtaining values
of the longitudinal magnetic field is standard and includes the following steps:
\par\noindent
1. The selection of lines suitable for measurement was carried out taking 
into account the physical properties and chemical composition of the star, 
calculating the synthetic spectrum with the STARSP code (Tsymbal 1996) 
using the VALD database (Kupka et al. 1999).
\par\noindent
2. in the observed spectral line profiles, corresponding to the 
left (LCP) and right (RCP) circular polarizations, we fitted a gaussian using
the method of least squares. Lines with defective 
registration due to the impact of cosmic rays, etc. were discarded. 
The value of line splitting under the effect of the magnetic field was 
determined by the magnitude of the shift between the centres of the fitted 
Gaussians.
\par\noindent
3. The intensity of the effective magnetic field for each line was computed 
using the well-known relation (Mathys 1991):

\begin{equation}
  \lambda_{R} - \lambda_{L} = 2 g_{\rm eff}\,\, 4.67 10^{-13} \lambda_0^{2} B_{\rm eff},
\end{equation}

\noindent
where the wavelength $\lambda$ is expressed in \AA, and the field
strength is in G. The value of $B_e$ was computed as the average over all the 
lines, used for measurements. Assuming that the scatter of $B_e$ values due to 
inaccurate measurements and other reasons is described by a normal law of 
distribution, the probable error $\sigma (B_e)$ was computed.
Instrumental effects are taken into account according to Bychkov et al. (2000).
During observations, we regularly registered the magnetic standards, $\alpha^2$
CVn and 53 Cam, and the standards of the null field $\alpha$ CMi, $\alpha$ Boo,
the Moon, etc., as described in Bychkov et al. (2006).
Results obtained for the standards are in good agreement with the literature data.

\subsection{Details of the literature data}

The full record of $B_e$ points analysed in this paper includes 372 CCD measurements
obtained from Zeeman splitting of metal lines, which were collected from published
papers. We list below references to the sources of previous data and the method of
observation. Fig.~\ref{fig:method} shows the distribution of $B_e$ points
obtained by different methods in the time period when magnetic measurements
were done (JD).

This is a very inhomogeneous set of observations certainly with underestimated
errors and possible offsets between various observers. The data set includes:

    \begin{figure}
    \resizebox{\hsize}{!}{\rotatebox{0}{\includegraphics{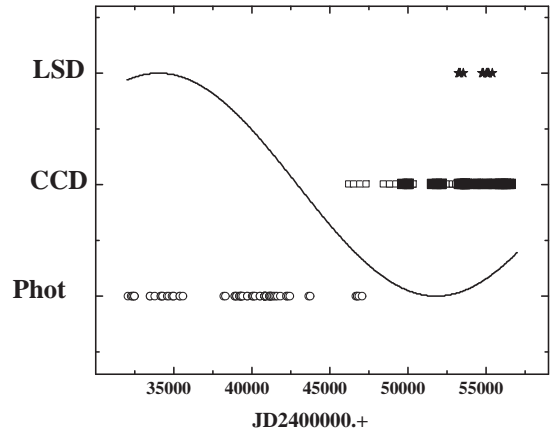}}}
    \vspace{0mm}
    \caption[]{Distribution of archival $B_e$ measurements (open circles)
       over years divided between particular methods: Phot - photographic
       method, CCD recording and LSD method. Solid
       line roughly indicates the run of $B_e$ with time. }
    \label{fig:method}
    \end{figure}

\begin{itemize}
\item Photographic estimates of $B_e$ taken from Babcock (1958), 
   Bonsack \& Pilachowski (1974), Scholz (1975; 1979), Zverko et al. 
   (1989) and Bychkov \& Shtol (1997).
\item CCD estimates (using CCD arrays and a more refined selection of 
   split lines). Open squares: measurements taken from Mathys (1991; 
   1994), Mathys \& Hubrig (1997), Hildebrandt, Scholz \& Lehmann (2000),
   Leone \& Kurtz (2003), Leone (2007) and Bychkov, Bychkova \& Madej
   (2006)  (filled squares)
\item LSD points (Least Squares Deconvolution method) are marked in 
   Fig.~\ref{fig:method} with asterisks. Source: Wade (2015).
\end{itemize}

\vspace{2mm} \noindent
Method (1) was used during the longest time period and covers phases of
the positive $B_e$ maximum on the magnetic phase curve of $\gamma$ Equ.

Method (2) required the use of echelle spectrometers to pack
many spectral orders on a CCD matrix of relatively small size. 
In fact CCD techniques use achromatic analysers of circular polarization
that operate in the range $\approx 2000 $ \AA (Babcock analyzed the interval
of only $300 - 350$ \AA). CCD analyzes higher number of spectral lines. 
The identification of lines is more accurate when using VALD and model 
atmospheres. In fact, this is just a more advanced version of method (1).

Method (3) is the most precise method. However, observations of the longitudinal
magnetic field of $\gamma$ Equ are located in a relatively short time
interval; they do not cover the minimum of the phase curve but are located at
the rising part of the curve. 

The principal point is that the local $B_e$ extrema were obtained mostly using
different methods. The $B_e$ maximum was observed by the photographic method and
the minimum by the CCD method (and LSD to a lesser extent). Differences and
possible offsets produced by these methods had an impact on the accuracy in
the determination of the period.

Note, that all of the $B_e$ estimates of Bychkov et al. (2006) and of this
work were obtained at the same telescope and spectrometer and processed 
with the same software. As can be seen from Fig.~\ref{fig:method}, these
estimates cover the minimum at the negative part of the magnetic phase
curve.

\subsection{Details of the LSD observations}

We appended to our time series 23 unpublished $B_e$ points obtained by the LSD
method (Wade 2015). These observations contributed to our period determination
with the same weight as other $B_e$ values.

These measurements were obtained from ESPaDOnS Stokes $V$ spectra
acquired between September 2004 and July 2010 as part of instrumental
commissioning and systematic monitoring of the instrumental
crosstalk. The spectra were reduced using Libre-Esprit (Donati et al.
1997) and LSD profiles were extracted using a line mask corresponding 
to an appropriate temperature and metallicity. Longitudinal field 
measurements were extracted from each LSD profile using the first 
moment method. Additional details concerning the reduction and analysis 
will be communicated in a forthcoming publication.

\subsection{Results}

The final $B_e$ time series extended for 24488 days (or over 67 years)
starting from the first measurement of the magnetic field on 9 October 
1946 by Babcock (1958). That time period amounts to ~69 percent of the 
expected long-term period of $\gamma$ Equ.

Belonging, as it does, to a rather late Ap spectral subclass, this star 
is a very good object for studying of global magnetic fields since it 
shows a large number of extremely narrow, sharp metal lines.

The available set of $B_e$ observations consists of 395 measurements 
obtained from Zeeman splitting of metal lines and 51 points measured
in hydrogen lines. 
Since values of $B_e$ (met) and $B_e$ (H line) are markedly different, 
we decided to determine periods using only estimates of $B_e$ (met). 
Moreover, the time interval over which $B_e$ (H line) values were observed
was 13131 days, which is approximately half the time interval for $B_e$ 
(met) observations, since the latter is equal to 24488 days.

The observed time series is very unevenly filled by estimates. For clarity, we constructed 
a histogram which shows the distribution of time intervals between neighbouring measurements.
Fig. 1 presents such a histogram with a step of 10 days. The horizontal axis in 
Fig. 1 was restricted to intervals not exceeding 350 days, but intervals
that are higher than 500 and 1000 days exist, and there is one interval equal
to 2614 days. 

    \begin{figure}
    \resizebox{\hsize}{!}{\rotatebox{0}{\includegraphics{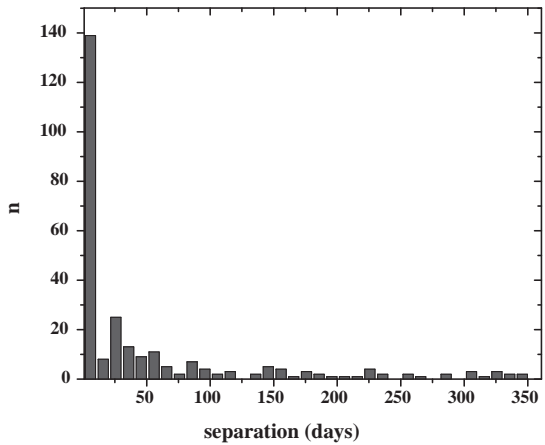}}}
    \vspace{0mm}
    \caption[]{ Distribution of time separation between two neighbouring
       $B_e$ measurements. }
    \label{fig:fig1}
    \end{figure}

\section{Methods of period determination}

There exist many different techniques for period determination [Lafler \& Kinman
(1965); Jurkevich (1971); Deeming (1975); Lomb (1976); Burki et al. (1978); Renson (1978);
Pelt (1975, 1983); Kurtz (1985); Scargle (1982) Terebizh (1992) and many others].
Those methods were transformed to usable computer codes or software packages
[e.g. Kurtz's code (1985) or the {\sc Period04} package (Goranskij 1976)]. 
One of the most serious problems in our case is that a given interval of the time
series data is shorter that the expected period. In the actual $B_e$ set it amounts only
to $\approx$ percent of the expected long-term period. Naturally the
accuracy of the period determination will be affected by unevenly distributed data
points of non-uniform accuracy. 

We selected the most commonly used methods and codes for the determination
of periods:
\begin{itemize}
\item 1. The Discrete Fourier Transform (DFT) for unevenly spaced data by
Deeming (1975), see also Scargle (1982) and Kurtz (1985). We used here
Fortran code kindly provided to us by Dr. Don Kurtz.
\item 2. Periodogram analysis (Burnasheva \& Gollandskij 1989). This is 
a modified version of Lomb (1976) and Scargle (1982).
\item 3. The least squares method. Determination of the minimum sum 
of squared deviations from the sine wave yields the best fit. The sum 
should be computed for a set of trial periods. The range of trial periods
and the frequency step are selected jointly with parameters of the analysed 
time series. The minimum sum indicates the most probable period. 
\end{itemize}

Method (3) was implemented in the form of the FORTRAN code
written by Bychkov (1988, unpublished). The advantage of this method is that
the user explicitly defines the shape of the phase curve which is a very
useful option in particular cases.

Each of the above most commonly used methods for period determination 
has some limitations of accuracy, especially when the length of the period
exceeds the length of time of the measurement set.

We also used a well-known package Period04 in the following tests
[Goranskij (1976), who implemented methods of Deeming (1975),
Lafler and Kinman (1965) and Jurkevich (1971)]. As indicated above, 
the specific case of $\gamma$ Equ is highly unusual, since the length 
of the $B_e$ time series equals 70 percent of the expected period.

\section{Analysis of observational data}

The simplest and most common form of phase curve is a simple harmonic curve

\begin{equation}
B_{e} (\phi) = B_0 + B_1 \sin (\phi) \, ,
\end{equation}
where
\begin{equation}
\phi = 2\pi \, \left( {{t - T_0} \over P} \right) \, .
\end{equation}

We performed a very simple numerical experiment. First, we estimated 
the preliminary value of the period $P_{\rm est}$ by fitting 
a sine wave to the observed $B_e$ points by least-squares (one degree 
of freedom). Then, for the fixed best period $P_{\rm est}=35394.22$ 
days, we determined the preliminary values of the half-amplitude $B_1=850.78$ 
G, the average $B_0=-266.24$ G and the initial epoch $T_0=2416365.12$ (JD).

Subsequently, we generated a new group of sine $B_e$ time series 
using the above preliminary estimates of parameters $B_0$, $B_1$,
$T_0$ and various trial periods $P_{\rm tr}$. Discrete strengths of 
the longitudinal magnetic field $B_e$ were determined at the same 
time points as in the observed time series. We then added a gaussian 
noise component with errors $\sigma$ in G given in specific source papers.

The trial periods $P_{\rm tr}$ of the simulated $B_e$ time series ranged from 
0.5 to 4 $\times$ $P_{\rm est}$. Then, for each $P_{\rm tr}$ we assessed 
again the period $P_{\rm re}$ using various methods and the relative
difference between them. In the ideal case, the trial period $P_{\rm tr}$
should be precisely reproduced.

\begin{figure}
    \resizebox{\hsize}{!}{\rotatebox{0}{\includegraphics{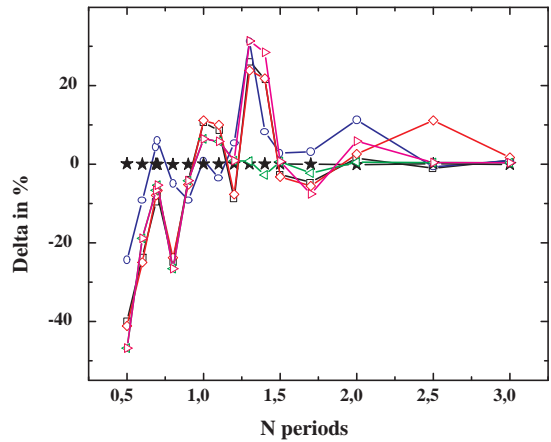}}}
    \vspace{0mm}
    \caption[]{Deviations of the re-estimated period $P_{\rm re}$ from 
       the trial value $P_{\rm tr}$ (in percent) as a function of the number 
       of variability cycles covered in the $B_e$ time series. Black 
       full asterisks -- results of preferred method (3).
       Grey line and open squares -- method (1) (Scargle); blue line and open circles
-- method (2) (program by Kurtz); red line and blank diamonds -- code Period04;
green line and open triangles -- code by Goransky (Lafleur-Kinman method);
pink line and open triangles -- code by Goransky (Deeming method). }
    \label{fig:p3c}
    \end{figure}

For clarity, we express the deviation R of the re-estimated period from the
trial value in percent, 
$R = (P_{\rm re} - P_{\rm tr})/ P_{\rm tr} \times 100 \% $.
The results of this simulation are shown in Fig.~\ref{fig:p3c} and 
Table~\ref{tab:fits}, which shows deviations $R$ as a function of the 
number of the trial periods contained in the specific $B_e$ time series.

For the length of the $B_e$ time series less than two expected periods, the most
reliable results are provided by method (3) (deviations $R$ close to zero). 
On the contrary, if the $B_e$ time  series is much longer than $P_{\rm tr}$
all cited methods are reliable (deviations $R$ converge to zero).

\begin{table}
\caption{Deviations of the re-estimated period $P_{\rm re}$ from the trial
   value $P_{\rm tr}$ (in percent) as a function of the number of trial periods
   covered by the $B_e$ time series (24487.3 days); see column 2. }
\label{tab:fits}
\begin{center}
\begin{tabular}{c r @{.} l  r @{.} l  r @{.} l  r @{.} l  r @{.} l }
\hline
P & \multicolumn{2}{c}{dt/P} & \multicolumn{2}{c}{R1} &
  \multicolumn{2}{c}{R2} & \multicolumn{2}{c}{R3} & \multicolumn{2}{c}{Per04}  \\
\hline
48974.60 &  &50 & -39&90 & -24&36 &  &12 & -41&18  \\
40812.20 &  &60 & -23&65 &  -9&23 & -&01 & -25&00  \\
35462.50 &  &69 &  -9&56 &   4&47 & -&03 &  -7&93  \\
34981.90 &  &70 &  -8&15 &   5&90 & -&04 &  -6&67  \\
30609.10 &  &80 & -24&88 &  -5&06 & -&13 & -23&81  \\
27208.10 &  &90 &  -4&05 &  -9&12 & -&05 &  -5&19  \\
24487.30 & 1&00 &  10&58 &    &98 &  &02 &  11&11  \\
22261.20 & 1&10 &   8&59 &  -3&55 &  &03 &  10&00  \\
20406.10 & 1&20 &  -8&83 &   5&44 &  &06 &  -7&69  \\
18836.40 & 1&30 &  26&14 &  31&28 &  &05 &  23&81  \\
17490.90 & 1&40 &  21&64 &   8&24 &  &03 &  21&83  \\
16324.90 & 1&50 &  -2&63 &   2&85 &  &04 &  -3&23  \\
14404.30 & 1&70 &  -4&55 &   3&17 &  &00 &  -5&56  \\
12243.70 & 2&00 &   1&53 &  11&26 & 2&56 &    &51  \\
 9794.90 & 2&50 &  -1&07 &   -&69 &  &41 &    &41  \\
 8162.40 & 3&00 &    &97 &    &91 & 1&70 &    &34  \\
\hline
\end{tabular}
\end{center}
\end{table}

The results of the period determination by the various methods are also 
shown in Fig.~\ref{fig:p3c}. Black full asterisks denote the preferred method
(3); gray line and open squares -- method (1) (Scargle); blue line and open circles
-- method (2) (program by Kurtz); red line and blank diamonds -- code Period04;
green line and open triangles -- code by Goransky (Lafleur-Kinman method);
pink line and open triangles -- code by Goransky (Deeming method). 

Fig.~\ref{fig:p3c} clearly shows that methods (1) and (2) yield reliable results only
when the length of time of a test series is not less than two or three periods. Method 
(3) is most independent of the duration of the time period. Of course, this simple
test gives only a rough idea of the performance of these methods when they
are applied to a specific time series. In fact, those features are useful
only to solve a specific problem of secular variability of the 
effective magnetic field of mCp star $\gamma$ Equ. 

Similar testing of various methods of period determination for 
simulated $B_e$ time series of length similar to the $\gamma$ Equ, both
equally and unevenly spaced, gave similar results.
Note, that this is a rare case, when the observed $B_e$ time series
is short, since the length of the time series is less than
the period of variations and this naturally affects the final result.

We finally accepted the value of the magnetic period obtained with method (3),
$P =35462.5$ days (97.16 years) -- as the most correct. It is evident that
using the very popular method (1) for this specific time series caused 
a reduction in the period of approximately 10 percent -- $P = 32051.69$ 
days (87.81 years), which is too low.

\section{Long term magnetic phase curve}

We obtained the following best least-squares fit to a long-term magnetic sine
wave for $\gamma$ Equ with the parameters:

\vspace{1.5mm} \hbox{\hskip12mm \vbox{
  \hbox{$T_0 = {\rm JD} \>\, 2416310.9 \pm 942 $ day }
  \hbox{$P = 35462.5 \pm 1148.6 $ days $= 97.16 \pm 3.15 $ year}
  \hbox{$B_0 = -265 \pm  6 $ G }
  \hbox{$B_1 = 850  \pm  8 $ G } } }

\begin{figure}
    \resizebox{\hsize}{!}{\rotatebox{0}{\includegraphics{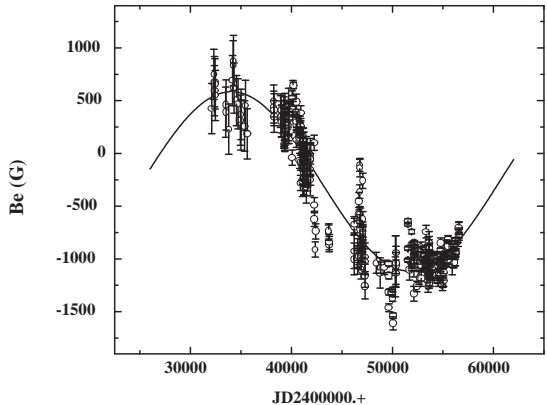}}}
    \vspace{0mm}
    \caption[]{ Full set of the available $B_e$ measurements versus time
       (in JD) for $\gamma$ Equ. Points $B_e$ in this plot were measured
       in metal lines. The solid line is the magnetic sine phase curve computed
       for the period 97.16 years. }
    \label{fig:metal2}
    \end{figure}

\noindent
The phase curve defined above is the best, final result of our research.
Fig.~\ref{fig:metal2} presents the phase curve and all of the $B_e$ points
drawn together (those derived from the splitting of metal lines).

Note, that the observed $B_e$ points exhibit a large and
non-gaussian scatter around the best-fitted sine wave. 
Such a scatter also can be enhanced by a rapid variability of the magnetic field,
comparable to rapid oscillations, for example, with periods in the
range of 6 to 30 minutes. See papers by Bychkov (1988), Leone \& Kurtz (2003),
Savanov et al. (2003), Hubrig et al. (2004) and Bychkov et al. (2005a). 

\section{Comparison with the most precise measurements }

\begin{figure}
    \resizebox{\hsize}{!}{\rotatebox{0}{\includegraphics{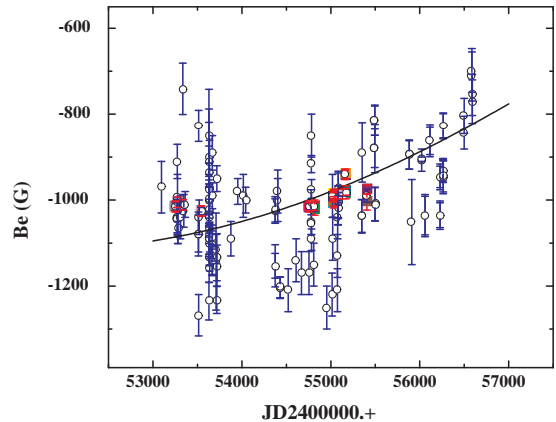}}}
    \vspace{0mm}
    \caption[]{ Section of our $B_e$ time series taken from Table 1,
       (blue points and error bars). Red squares denote LSD $B_e$ points
       of years 2004-2012 (Wade 2015) compared to our best-fit
       sine wave. Note that the LSD $B_e$ points are located well between
       our points of lower accuracy and are close to the phase curve. }
    \label{fig:l2}
    \end{figure}

In the period and phase curve determination, we used 23 recent measurements 
of the longitudinal magnetic field $B_e$ of $\gamma $ Equ measured in
metal lines with the precise LSD method (Wade 2015).
Fig.~\ref{fig:l2} shows our homogeneous $B_e$ measurements collected in
Table 1 (blue circles with error bars) and the phase curve in JD just
after the $B_e$ minimum. LSD measurements are concentrated in the area of
red empty squares close to the phase curve.

LSD measurements alone essentially form an almost horizontal line and then 
suggest that real period of $\gamma$ Equ is much longer than $\approx$
100 years, in contrast to our current determination (Wade 2015).
We note, that such a hypothesis should be verified by further LSD observations
performed using the same instrument, since the current LSD data cover only a
small period of time and such a low rise in $B_e$ tells us little about 
the long term behaviour of the LSD longitudinal magnetic field. 

We believe, that precise LSD measurements of the longitudinal magnetic
field (those obtained from the splitting of metal lines) are consistent with
the new phase curve and our other conclusions. 

\begin{figure}
    \resizebox{\hsize}{!}{\rotatebox{0}{\includegraphics{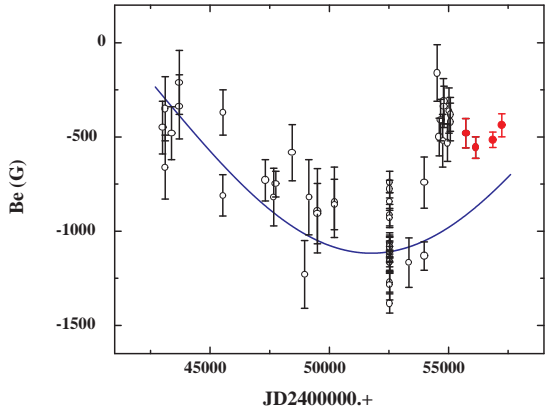}}}
    \vspace{0mm}
    \caption[]{ All of the available $B_e$ measurements obtained from hydrogen
       lines plotted against JD. Red points denote recent measurements of
       the $B_e$ field obtained using the DAO dimaPol method (Bohlender 2015).
       The solid line is our final $B_e$ phase derived from metal lines.
       Note the apparent offset of $B_e$ values measured in hydrogen lines.}
    \label{fig:bagh2}
    \end{figure}

Additionally, Fig.~\ref{fig:bagh2} shows the run of all available $B_e$ measurements
for $\gamma$ Equ, which were obtained from hydrogen lines (51 points). Red
dots denote recent DAO dimaPol determinations of $B_e$ in $\gamma$ Equ
[(Bohlender (2015); see Monin et al. (2012) for a description of that technique].
Note, that the $B_e$ values obtained from hydrogen lines
exhibit an offset compared to those from metal lines after the minimum.

We did not use the latter $B_e$ points for period determination because
of that offset, which is a fact that has been well-known from a long time.
Hydrogen is uniformly distributed on the surface of stars and metals apparently
are not. Minor deviations from the apparently uniform
distribution of HI can be observed only at the He-rich and He-weak stars
(Kudryavtsev \& Romanyuk 2012.

Moreover, the time interval covered by the hydrogen line estimates is only 13131
days (37 percent of the period) and the corresponding $B_e$ time series contains
only 51 points. 

\section{Summary}

In this paper, we present a series of 95 new measurements of the longitudinal 
magnetic field $B_e$ of Ap star $\gamma$ Equ = HD 201601, which were
obtained over 4190 days (more than 11 years) at the Special Astrophysical Observatory.
We also compiled and analysed all the published and available unpublished
LSD $B_e$ data for this star.
A useful subset of magnetic data consists of 395 $B_e$ measurements 
obtained in Zeeman split metal lines over 24488 days (more than 67 years).

We examined the reliability of the most frequently applied methods for period
determination in the rare case, in which the length of the observed $B_e$ time
series is lower than an unknown period (in the case of $\gamma$ Equ it equals
$\approx 69$ percent of the period $P_{\rm mag}$). In such a case the most 
reliable technique is the least squares method of fitting a sine wave to 
the actual $B_e$ time series. However, such a conclusion applies only to
the particular record of $B_e$ observations of $\gamma$ Equ analysed in
this paper.

\begin{enumerate}
\item 1. In the uncommon case of a relatively short time series of $B_e$
measurements which is shorter than an unknown period, the fitting of a sine 
wave to the time series is the only reliable method for determination of
the magnetic (rotational) period. It is possible, that such a conclusion 
is restricted to the specific set of $B_e$ measurements of $\gamma$ Equ.
\item 2. We determined that the period of secular variations $P=35462.5$
days (97.16 years) and is longer than the previously accepted value.
\item 3. We present a series of 23 precise LSD measurements of its longitudinal
magnetic field (Wade 2015) which confirms the accuracy of our measurements.
\end{enumerate}

\section*{Acknowledgments}

The authors sincerely thank Don Kurtz for providing the Fortran code 
which was used in this paper for analysing the time series.
We thank David Bohlender, the referee and Gregg Wade 
for extensive discussion and criticism of our results. 
This research was supported by Polish National Science 
Center grant No. 2011/03/B/ST9/03281 and Russian grants ``Leading Scientific
Schools'' N2043.2014.2. and President grants MK-6686.2013.2 and MK-1699.2014.2.  
Research by Bychkov V.D. was supported by the Russian Scientific Foundation
grant N14-50-00043.

\clearpage

\section{appendix A}
\begin{table}
\caption{Magnetic field measurements, $B_e$, of the Ap star $\gamma$
    Equ, obtained with the 1-m telescope at SAO RAS. }
\label{tab:saores}
\renewcommand{\arraystretch}{1.1}
\begin{center}
\begin{tabular}{|c|r|r|c|r|r|}
\noalign{\vskip 2 mm}
\hline
JD2400000.+ $\hskip-2mm$ & $B_{e}$ & $\hskip-1mm$ $\sigma_{B_e}$ &$\hskip2mm$
JD2400000.+ $\hskip-2mm$ & $B_{e}$ & $\hskip-1mm$ $\sigma_{B_e}$ \\
\hline
53274.38 & -1041. &  35. &  54430.16 & -1203.  &  24. \\  [-1.0pt]
53274.40 & -1041. &  38. &  54779.25 & -1005.  &  29. \\  [-1.0pt]
53275.43 &  -995. &  55. &  54780.14 & -1052.  &  29. \\  [-1.0pt]
53277.43 & -1042. &  44. &  54780.18 & -1053.  &  27. \\  [-1.0pt]
53278.41 & -1043. &  38. &  54781.14 &  -916.  &  20. \\  [-1.0pt]
53278.43 & -1042. &  41. &  54782.15 &  -974.  &  22. \\  [-1.0pt]
53279.37 & -1064. &  37. &  54783.17 & -1092.  &  25. \\  [-1.0pt]
53279.39 & -1064. &  37. &  55081.32 & -1020.  &  37. \\  [-1.0pt]
53346.25 & -1025. &  38. &  55081.36 & -1020.  &  38. \\  [-1.0pt]
53511.47 & -1078. &  43. &  55082.31 & -1004.  &  37. \\  [-1.0pt]
53511.51 & -1079. &  51. &  55082.34 & -1005.  &  38. \\  [-1.0pt]
53513.52 & -1268. &  48. &  55083.28 &  -971.  &  38. \\  [-1.0pt]
53514.53 &  -827. &  36. &  55083.31 &  -971.  &  34. \\  [-1.0pt]
53628.35 & -1235. &  44. &  55348.48 & -1037.  &  38. \\  [-1.0pt]
53629.27 &  -902. & 160. &  55348.50 & -1038.  &  39. \\  [-1.0pt]
53629.29 &  -973. &  99. &  55351.52 &  -889.  &  69. \\  [-1.0pt]
53629.31 & -1001. &  61. &  55494.17 &  -815.  &  34. \\  [-1.0pt]
53629.34 & -1028. &  58. &  55494.20 &  -814.  &  35. \\  [-1.0pt]
53629.36 &  -851. &  63. &  55495.18 &  -879.  &  45. \\  [-1.0pt]
53629.38 &  -974. &  60. &  55495.21 &  -880.  &  56. \\  [-1.0pt]
53629.40 & -1023. &  54. &  55496.18 & -1009.  &  39. \\  [-1.0pt]
53629.43 & -1065. &  67. &  55496.21 & -1010.  &  39. \\  [-1.0pt]
53629.45 & -1040. &  87. &  55881.27 &  -894.  &  34. \\  [-1.0pt]
53629.47 &  -937. & 118. &  55881.30 &  -895.  &  33. \\  [-1.0pt]
53630.25 & -1002. &  40. &  55911.12 & -1051.  &  99. \\  [-1.0pt]
53630.27 & -1002. &  57. &  56021.55 &  -907.  &  26. \\  [-1.0pt]
53632.36 & -1159. &  33. &  56021.61 &  -905.  &  25. \\  [-1.0pt]
53634.30 & -1084. &  52. &  56058.53 & -1036.  &  49. \\  [-1.0pt]
53635.27 & -1102. &  51. &  56058.55 & -1036.  &  45. \\  [-1.0pt]
53635.29 & -1102. &  49. &  56114.47 &  -862.  &  37. \\  [-1.0pt]
53636.25 &  -912. &  51. &  56114.49 &  -862.  &  32. \\  [-1.0pt]
53637.23 & -1133. &  32. &  56228.26 &  -946.  &  25. \\  [-1.0pt]
53637.25 &  -997. &  29. &  56228.29 &  -946.  &  30. \\  [-1.0pt]
53638.21 & -1091. &  35. &  56229.20 & -1035.  &  29. \\  [-1.0pt]
53638.23 & -1090. &  31. &  56229.23 & -1035.  &  33. \\  [-1.0pt]
53666.17 & -1104. &  32. &  56262.12 &  -931.  &  25. \\  [-1.0pt]
53666.19 & -1104. &  38. &  56262.14 &  -931.  &  28. \\  [-1.0pt]
53667.18 & -1141. &  32. &  56263.11 &  -827.  &  28. \\  [-1.0pt]
53667.21 & -1140. &  36. &  56263.14 &  -827.  &  30. \\  [-1.0pt]
53668.19 & -1086. &  28. &  56264.12 &  -946.  &  39. \\  [-1.0pt]
53718.12 & -1232. &  32. &  56264.14 &  -945.  &  36. \\  [-1.0pt]
53719.11 & -1133. &  55. &  56500.40 &  -843.  &  38. \\  [-1.0pt]
53721.12 & -1153. &  48. &  56500.43 &  -804.  &  41. \\  [-1.0pt]
54373.19 & -1024. &  27. &  56587.18 &  -712.  &  57. \\  [-1.0pt]
54373.22 & -1023. &  31. &  56587.21 &  -698.  &  51. \\  [-1.0pt]
54374.20 & -1153. &  29. &  56590.15 &  -755.  &  48. \\  [-1.0pt]
54374.23 & -1153. &  49. &  56590.18 &  -771.  &  51. \\  [-1.0pt]
54430.13 & -1204. &  25. &           &         &      \\  [-1.0pt]
\hline
\end{tabular}
\end{center}
\end{table}


\begin{thebibliography}{}

\bibitem{} Babcock, H.W. 1958, Ap\&SS, 3, 141
\bibitem{} Bohlender, D. 2015, private communication
\bibitem{} Bonsack, W.K., Pilachowski, C.A. 1974, ApJ, 190, 327
\bibitem{} Burki, G., Maeder, A., Rufener, F. 1978, A\&A, 65, 363
\bibitem{} Burnasheva, B.A., Gollandskij, O.P. 1989, BCrAO, 77, 177
\bibitem{} Bychkov, V.D., 1988, Proc. Int. Meet. on :
   "Physics and Evolution of Stars ", held in Nizny Arkhyz, October
   12-17, 1987. Edited by Yu.V.Glagolevsky. Leningrad: Nauka, 197
\bibitem{} Bychkov, V.D., 2008, Astrophysical Bulletin, 63, 83
\bibitem{} Bychkov, V.D., Shtol', V.G. 1997, Stellar Magnetic Fields, Proc. Int. Conf., 200
\bibitem{} Bychkov, V.D., Bychkova, L.V., Madej, J. 2005a, Acta Astronomica, 55, 141
\bibitem{} Bychkov, V.D., Bychkova, L.V., Madej, J. 2005b, A\&A, 430, 1143
\bibitem{} Bychkov, V.D., Bychkova, L.V., Madej, J. 2006, MNRAS, 365, 585
\bibitem{} Bychkov, V.D., Romanenko, V.P., Bychkova, L.V., 2000, Bull. Spec. Astrophys. Obs., 49, 147
\bibitem{} Deeming, T.J. 1975, Ap\&SS, 36, 137
\bibitem{} Donati, J.-F., Semel, M., Carter, B.D., Rees, D.E., Cameron, A.C., 1997, MNRAS, 291, 658
\bibitem{} Goranskij, V.P. 1976, PZP, 2, 323
\bibitem{} Hildebrandt, G., Scholz, G., Lehmann, H. 2000, AN, 321, 115
\bibitem{} Hubrig, S., Kurtz, D.W., Bagnulo, S., Szeifert, T., Scholler, M.,
    Mathys, G., Dziembowski, W.A. 2004, A\&A, 415, 661
\bibitem{} Jurkevich, I. 1971, Ap\&SS, 13, 154
\bibitem{} Kudryavtsev, D.O., Romanyuk, I.I. 2012, AN, 333, 41
\bibitem{} Kupka, F., Piskunov, N.E., Ryabchikova, T.A. et al., 1999, A\&AS, 138, 119
\bibitem{} Kurtz, D.W. 1985, MNRAS, 213, 773
\bibitem{} Lafler, J., Kinman, T.D. 1965, Ap\&SS, 11, 216
\bibitem{} Leone, F. 2007, MNRAS, 382, 1690
\bibitem{} Leone, F., Kurtz, D. 2003, A\&A, 407, L67
\bibitem{} Leroy, J.L., Bagnulo, S., Landolfi, M., Degli'Innocenti, E. Landi, 1994, A\&A, 284, 174
\bibitem{} Lomb, N.R. 1976, Ap\&SS, 39, 447
\bibitem{} Mathys, G. 1991, A\&ASS, 89, 121
\bibitem{} Mathys, G. 1994, A\&ASS, 108, 547
\bibitem{} Mathys, G., Hubrig, S. 1997, A\&ASS, 124, 475
\bibitem{} Monin, D., Bohlender, D., Hardy, T., Saddlemyer, L., Fletcher, M. 2012, PASP, 124, 329
\bibitem{} Pelt, J. 1975, Tartu Astrofuus. Obs. Teated, 52, 24
\bibitem{} Pelt, J. 1983, in ESA Statist. Methods in Astron., 37
\bibitem{} Renson, P. 1978, A\&A, 63, 125
\bibitem{} Savanov, I., Musaev, F.A., Bondar, A.V. 2003, IBVS, 5468
\bibitem{} Savanov, I.S., Romanyuk, I.I., Semenko, E.A., Dmitrienko, E.S. 2014, psce.conf, 386
\bibitem{} Scargle, J.D. 1982, ApJ, 263, 835
\bibitem{} Scholz, G. 1975, AN, 296, 31
\bibitem{} Scholz, G. 1979, AN, 300, 213
\bibitem{} Scholz, G., Hildebrandt, G., Lehmann, H., Glagolevskij, Yu.V. 1997, A\&A, 325, 529
\bibitem{} Schuster, A. 1898, Terrestrial Magnetism (now GJR), 3, 13
\bibitem{} Terebizh, V.Yu. 1992, Time Series Analysis in Astrophysics, Moscow, Nauka (in Russian).
\bibitem{} Tsymbal, V. 1996, ASP Conference Series, 108, 198
\bibitem{} Wade, G.W. 2015, private communication
\bibitem{} Zverko, J., Bychkov, V.D., Ziznovsky, J., Hric, L. 1989, 
             Contr. Astron. Obs., Skalnate Pleso, 18, 71
 
\end{thebibliography}
\end{document}